# Catalytic Nucleation of Amyloid Beta and Hen Egg White Fibrils, and p53 Oligomerization

J. C. Phillips

Dept. of Physics and Astronomy, Rutgers University, Piscataway, N. J., 08854

Abstract

Scaling theory generates transferable (even universal) algebraic and geometrical relations between the amino acid sequences and the aggregation functions of the three titled radically different proteins. In addition to the two hydropathicity scales and β strand scales used in earlier p53 work, a second β strand Hot Spot scale is shown to yield very accurate results for oligomerization of p53, the "tumor suppressor". These algebraic and geometrical relations could be caused topologically by the dominance of protein-protein aggregation by interactions in a membrane catalytic surface layer.

**1. Introduction**

β–Amyloid proteins Aβ are 40-42 amino acid (aa) fragments of the 770 aa Amyloid precursor protein A4, which has a very large 80 aa hydrophobic peak between 672 and 752 [1]. This C-terminal peak has spinodal thermodynamic features: it splits symmetrically in half, with the lower 40-42 aa fragment forming Aβ [2]. Aβ proteins form antiparallel β sheets which are strongly amphiphilic [3], and which aggregate to form toxic oxidative Aβ fibrils [3-5]. Similar β aggregations occur often, and have been studied in >13,000 articles. General models of these structures are based on cross-β spines similar to the fibril-forming segment NNQQNY of the yeast prion protein Sup35 [6]. Successive parallel β sheets are closely fitted (interdigitated in velcro-like steric zippers) [7]. Many examples are known experimentally, although most examples have been obtained in extreme conditions.

Aβ fibrils themselves are composed of amphiphilic 13-mer modular β sandwiches separated by reverse turns [8]. Hydropathic waves were used to optimize the description of the small (40,42 aa) plaque-forming (aggregative) Aβ fragments, resulting in successful predictions of edge and center features of the fragments [2]. The hydropathic wave method involves uses only one adjustable parameter, the hydropathic wave length W, to obtain the profile ψ(aa,W). W is the width of a sliding window, which is used to smooth profiles obtained from a suitable bioscale ψ(aa,1).

So far only the present author has tested different bioscales ψ(aa,1) to determine their effectiveness for various protein properties, including amyloid formation and evolution of multiple functions. Previous studies have always fixed W at a small value (usually 5); only the present author has varied W in order to optimize its value in each context. The comparative advantage of using multiple bioscales and multiple length scales is that correlations between sequences and function can be optimized and become operationally obvious.

The modern MZ hydropathicity scale [9] for ψ(aa,1) is based on bioinformatic studies of the differential solvent-exposed surface geometry of thousands of PDB segments spanning a length range centered on 20 aa. It has given impressively superior results for large-scale properties of many proteins, compared to the classic KD scale [10], based on water-air enthalpy differences of small synthetic peptides.

Because β aggregations occur often, one should consider the bioinformatic FTI scale of exposed β strand propensities, based again on surveys of thousands of PDB structures [11]. A little-known Hot Spot (βHS) scale is based on direct measurement of amyloid formation by a 7 aa Aβ nucleus at the center of Aβ [12]. It turns out that the MZ, KD and FTI scales are strongly correlated, with r ≈ +/−0.9. However, the MZ scale with W = 13 is superior in recognizing the 13-mer Aβ sandwich structure [2,8]. With W = 21 both the KD and MZ scales successfully identify the Aβ hydrophilic cleavage dip at Asp672, and end at the hydrophobic peak 712 ± 1. The KD scale does slightly better at identifying the similar stabilities of the 40- and 42-aa Aβ fragments. Interestingly, r(KD,FTI) = -0.93, while r(MZ,FTI) = -0.91, so the KD scale includes some extra β strand features. We will find interesting correlations between the universal FTI βexp scale and the βHS scale. The values of the two β scales used here are in Table I.

The choice W = 21 is also natural, because 21 aa is the typical membrane thickness for transmembrane proteins [13], and it is the center of the segmental fractal range 9-35 fitted by MZ [9]. One can carry this description to the similarly 20 amino acid thick "frontier" layers adjacent to cell membranes where proteins can interact most effectively, as they are temporarily confined to a narrower space. Biochemists may recognize a similarity to heterogeneous catalysis, with the membrane here serving as catalytic substrate. This picture is a generalization of enzyme catalysis (proteins with organic molecules) to all protein-protein interactions and evolution in vivo.

With the choice W = 21, the L688 − I712 of A4 (L17-I42 of Aβ) region derived from NMR data [8] becomes steeply and nearly linearly hydrophilic (Fig. 1). We can picture Aβ (1-40,42) in the frontier layer as orienting its hydrophobic maximum near 712 adjacent to the membrane, while its hydrophilic minimum is at the edge of the frontier layer.

**2. Aβ Nuclei**

While W = 21 gives the best description of amphiphilic Aβ 1-40 (42), much shorter hydrophobic segments are often best described with smaller values of W. One of the shortest and best studied Aβ 40-42 aa segments is central 16-22 (687-693, KLVFFAE) [12,14]. This 7 aa segment is best described with W = 9, as shown in Fig. 2. W = 9 also gives a good fit to the 7 aa Aβ segment 31-37 (702- 708, IIGLMVG), see Fig. 4 (c) of [7]. The experimental 23-30 gap between these two heptamers is a false positive for the parallel β sheet Velcro method [7], but it is correctly predicted here with W = 9. It is also correctly predicted by the HS scale, as shown in Fig.3.

Because the hydropathic scales MZ and KD are strongly anticorrelated to the FTI βexposed and HS scales, a plot (not shown) similar to Fig. 2, but using the FTI βexposed scale also successfully separates the 16-22 and 31-37 Aβ nuclei, which now appear as βexposed minima. However, the contrast between the 21 and 27-28 extrema is 30% smaller with the βexposed scale.

A subtle point, which illustrates the importance of the connection between length scale and functionality, concerns the 709-712 C-terminal section in Fig. 3. This shows a peak in the FTI βexp and βHS, W = 9, profiles (both emphasizing β strand propensities), but not in the hydropathic KD and MZ profiles. According to [12], the 709-712 section should be counted as an amyloid nucleus, because Aβ 42 (670-712) is a stronger amyloid former than Aβ 40 (670-710). However, this comparison involves entire Aβ fragments W ≈ 40, whereas the smaller nuclei involve W ≈ 9. The larger scale erases the 23-30 gap between the two small 687-693 and 702-708 nuclei, and also spreads the latter over into the 709 - 712 section.

The correlation between the r(HS, FTI) W = 1 β scales is only -0.87, but over the wide C terminal range 670 – 730 of Fig. 3 it is -0.96. Thus the precursor protein A4 is stabilized against amyloid formation by a deep spinodal valley in βexposed propensity [1]. The mutational method of [12] has revealed this correlation directly, as one can see from Fig. 3 and from this very strong correlation. This is a remarkable success for mutagenics, which is made possible by choosing the center of central 16-22 (687-693) as the mutagenic site for βHS.

## 3. Globular Lysozyme Nuclei

The aggregation of globular proteins, such as well-studied lysozyme *c* (Hen Egg White), may involve unfolding, and is thus more complex than that of Aβ, a known product of A4 fragmentation. The smallest lysozyme amyloid nucleus is 55-63 (9 aa) GIFQINSRY, called K peptide [15,16], renumbered as 73-81 in Uniprot P61626. K peptide is the strongest amyloid former of nine related small (< 9 aa) peptides over a pH range from 2 to 9. Profiles for entire human lysozyme in Fig. 4 show that the 9 aa K peptide nucleus is located at the center of centrosymmetric α-β-α lysozyme. The 69 aa wide central β region 57-103 is hydropathically level, so its β strands are nonamphiphilic [17].

Folding is a global property, and the β central region (57-103) of lysozyme *c* dominates its evolution from Chicken to Human, as its average hydropathicity increases (Fig. 2 of [17]) and its globular roughness peak at W = 69 decreases (Fig. 1 of [17]). The best scale here is the long-range MZ scale, which gives the Chicken/Human roughness ratio at W = 69 as 3.0. The short-range scales KD, βexp, and βHS, have respective ratios of 1.9, 1.3 and 1.8. Thus at long wave lengths, the long-range MZ hydropathic scale has twice the resolving power of the three short-range scales. Even so, the superiority of the βHS scale to the βexp scale in describing HEW evolution in the β central region (57-103) is noteworthy. It is explained by scaling, because βHS is associated with W = 9, whereas the construction of the βexp scale involves W ~ 1.

## 4. Non-Globular p53

Most proteins are hydropathically compacted into globules, but the tumor suppressor p53 forms a flexible, tetrameric four-armed starfish [18], quite distinct from the globular structures which most proteins (even when oligomerized) form. Among all proteins p53 is much more hydrophilic than average, and it is also elastically much softer, with about half its structure dominated by β strands [18], while the remainder (especially the N-terminal quarter) is disordered. Previously fits to the epitopes used for early cancer detection through binding to autoantibodies [19] were examined for the two hydropathic scales MZ and KD, and the βexp scale, with the latter being the most effective [18].

Given the improved recognition of the aggregative β central region (57-103) of HEW in Fig. 5 by the βHS scale compared to the βexp scale, one optimistically asks whether the βHS scale could be better in resolving p53 epitopes. Instead, the epitopic W= 9 βHS p53 profiles very closely resemble (r = 0.95) the ineffective W = 9 MZ profiles, while other profile pairs are quite distinct.

This correlation, of two scales so different in origin, can be exploited to quantify other aspects of p53 functionality. First we should determine the length scales L associated with the MZ – βHS W = 9 correlation (Fig. 6). The correlations (usually > 0.9) are so strong that several features of r(9,L) averaged over all p53 are apparent (figure caption). These can be profiled with as small a length scale as W = 9 and L = 1, because 9 is the lower limit of the fractal segments studied by MZ [9].

The p53 family proteins have a modular domain organization comprising DNA-binding and tetramerization domains (TD) that are linked and flanked by intrinsically disordered regions with high sequence diversity, which have been analyzed both by multiple sequence alignment and for their secondary structure [20]. We look first at the MZ – βHS W = 9 correlation in TD 325-355. This is very high for humans, r = 0.985, while for chickens 325-355 is only r = 0.897 (~ 0.9, ordinary scale correlation).

To reduce this difference, we can narrow the TD by 13% to the clustal range suggested by multiple sequence alignment, 326-352 [20]. Here the human (chicken) correlations are 0.978 (0.928). The difference is reduced, both because of a decrease in r(9,hum) and an increase in r(9,chk). Is the choice W = 9 best? It is: with W = 7 for humans, r = 0.975, and W = 11 gives 0.969.

These precise results confirm the effectiveness of clustal analysis. They also show how nonlocal scaling can reveal evolutionary oligomer improvements not accessible to clustal and secondary α helix analysis [20]. The human TD r(MZ – βHS  W = 9) has evolved to be nearly complete (r ~ 1), and much larger than the chicken TD r, providing additional oligomer stability through wider and stronger synergies of short range HS interactions, and long range (strain) hydropathic interactions.  This leads us immediately to the next binding question, that of p53 epitopes to autoantibody paratopes [18], which will be discussed elsewhere.

## 5. Discussion

The mechanism of aggregation of protein fibrils based on formation of parallel β sheets was discussed in [7] in detail for Aβ, lysozyme, myoglobin and tau.  This special packing yielded largely positive results, even though Aβ forms antiparallel β sheets and lysozyme is a mixture of α helices and β strands, while myoglobin has only α helices, as does tau.  There were also false positives, whose number increased from Aβ to the other proteins containing α helices (Fig. 4 of [7]).  The positive successes of [7] can be understood as the result of amphiphilically driven conformational flexibility [20,21].

A much more elaborate method for predicting amyloidal aggregation involves multiple variables (typically seven) scanned over a sliding window width fixed at 7 [22,23], and adjustment of corresponding (typically seven) weighting parameters.

Given the exponentially complex nature of protein amino acid packing, even the limited success of the parallel β sheet model is most impressive.  By contrast, there appears to be no packing geometry involved in hydropathic wave scaling.  How is it possible for scaling to be more successful in describing anisotropic 3D fibril formation when it contains no obvious geometry?  Evolution has built many universal geometrical features into the hydropathic KD and MZ scales, as well as the βexpos FTI and βHS scales, as proved by their strong correlations (~0.9), although each scale is derived in a different way.  These correlations are the biological realization of what computer scientists are calling "deep neural networks" [24].

To utilize these scales effectively to quantify aggregation properties, one must construct scaling waves ψ(aa,W), and select the value of W best suited to the initial nucleation stages of  fibril aggregation.  The choice of W is usually a natural one, but in any case the result is usually easily justified thermodynamically.  Here, for example, the shorter KD scale has proved most effective for small 7 aa nuclei, because it was based originally on small synthetic peptides.  Similarly, the MZ scale is most accurate for 40 aa Aβ itself, because it was derived from protein segments with lengths centered on 21 aa. We have shown that long-range interactions, described by the MZ scale [9], correlate synergistically with short-range interactions, described by the βHS scale [12], to stabilize the TD of p53.

Protein expression using Escherichia coli is a common and important method for recombinant protein production. There is a strong positive relation between the thermodynamic stability of p53 tetramers against single mutants in and near the TD and TD expression in *E. coli* [25]. The correlations discussed here are also a powerful new principle for quantifying the binding of p53 epitopes to autoantibody paratopes [18]. These epitopes are the most promising noninvasive molecular biomarkers that can detect and diagnose cancers in a cost-effective manner at an early stage. Such a blood–based test would be the Holy Grail of both cancer treatment and combination chemotherapy [26], and will be discussed further elsewhere.

In Table I the values of Ψ(aa) are conveniently scaled to span the same range as the $\Psi_{MZ}$ values in [27]. Surprisingly the scaling for βHS [12] requires only a multiplicative factor, without an additive constant. This means that amyloid formation is driven by the 18-26 Aβ nucleus center aa, and is almost proportional to the MZ fractal defined over a much larger and wide range of solvent exposed areas averaged over all proteins. Even granted the critical importance of this Aβ nucleus to the stability of the neural network, this universal relation is very surprising. It is plausible that these correlations are possible because protein aggregation is dominated by interactions in a membrane catalytic surface layer.

**Methods** The calculations described here are very simple, and are most easily done on an EXCEL macro. The one used in this paper was built by Niels Voorhoeve and refined by Douglass C. Allan.

# References

1. J. C Phillips, A cubic equation of state for amyloid plaque formation. arXiv1308.5718 (2013).

2. J. C. Phillips Thermodynamic description of beta amyloid formation using physicochemical scales and fractal bioinformatic scales. ACS Chem. Neurosci. **6**, 745-750 (2015).

3. D. Schubert, C. Behl, R. Lesley, A. Brack, R. Dargusch, Y. Sagara, H. Kimura, Amyloid peptides are toxic via a common oxidative mechanism. Proc. Nat. Acad. Sci. (USA) **92**, 1989-1993 (1995).

4. C. Behl (1999) Alzheimer's disease and oxidative stress: Implications for novel therapeutic approaches. Prog. Neurobio. **57**, 301-323.

5. F. Souza, E. Michelli, A. S. de Vasconcelos, T. Costa Vilhena, et al. Oxidative Stress in Alzheimer's Disease: Should We Keep Trying Antioxidant Therapies? Cell. Mol. Neurobiol. **35**, 595-614 (2015).

6. M. R. Sawaya, S. Sambashivan, R. Nelson, ; et al. Atomic structures of amyloid cross-beta spines reveal varied steric zippers. Nature **447**, 453- 457 (2007).

7. M. J. Thompson, S. A. Sievers, J. Karanicolas, et al. The 3D profile method for identifying fibril-forming segments of proteins. Proc. Nat. Acad. Sci. (USA) **103**, 4074-4078 (2006).

8. T. Luhrs, C. Ritter, M. Adrian, D. Riek-Loher, B. Bohrmann, H. Doeli, D. Schubert, R. Riek 3D structure of Alzheimer's amyloid-beta(1-42) fibrils. Proc. Nat. Acad. Sci. (USA) **102**, 17342-1734 (2005).

9. M. A. Moret, G. F. Zebende, Amino acid hydrophobicity and accessible surface area. Phys. Rev. E **75**, 011920 (2007).

10. J. Kyte , R. F. Doolittle, A simple method for displaying the hydropathic character of a protein**.** J. Mol. Biol. **157,** 105-132 (1982).

11. K. Fujiwara, H. Toda, M. Ikeguchi, Dependence of alpha-helical and beta-strand amino acid propensities on the overall protein fold type. BMC Struc. Biol. **12**, 18 (2012).

12. N. S. de Groot, I. Pallares, F. X. Aviles, et al. Prediction of "hot spots" of aggregation in disease-linked polypeptides. BMC Struc. Biol. **5**, 18 (2005).

13. J. C. Phillips, (2010) Self-organized criticality in proteins: hydropathic roughening profiles of G-Protein Coupled Receptors. Phys. Rev. E **87**, 032709 (2013).

14. J. J. Balbach, Y. Ishii, O. N. Antzutkin, et al., Amyloid fibril formation by A beta(16-22), a seven-residue fragment of the Alzheimer's beta-amyloid peptide, and structural characterization by solid state NMRBiochemistry **39,** 13748-13759 (2000).

15. Y. Sugimoto, Y. Kamada, Y. Tokunaga, et al. Aggregates with lysozyme and ovalbumin show features of amyloid-like fibrils. Biochem. Cell Biol. **89**, 533-544 (2011).

16. Y. Tokunaga,; M. Matsumoto, Y. Sugimoto, Amyloid fibril formation from a 9 amino acid peptide, 55th-63rd residues of human lysozyme. Int, J. Bio. Macromol. **80**, 208-216 (2015).

17. J. C. Phillips, Fractals and Self-Organized Criticality in Proteins. Phys A **415**, 440-448 (2014).

18. J. C. Phillips, Autoantibody recognition mechanisms of p53 epitopes. Physica A **451**, 162-170 (2016).

19. J. W. Pedersen, A. Gentry-Maharaj, E-O. Fourkala, et al. Early detection of cancer in the general population: a blinded case-control study of p53 autoantibodies in colorectal cancer. British J. Cancer **108**, 107-114 (2013).

20. A. C. Joerger, R. Wilcken, A. Andreeval, Tracing the Evolution of the p53 Tetramerization Domain. Structure **22**, 1301-1310 (2014).

21. D. J. Gordon, J. J. Balbach, R. Tycko, et al. Increasing the amphiphilicity of an amyloidogenic peptide changes the beta-sheet structure in the fibrils from antiparallel to parallel. Biophys. J. **86**, 428-434 (2004).

22. E. Calabro, S. Magazu, Transition from alpha-helix to beta-sheet structures occurs in myoglobin in deuterium oxide solution under exposure to microwaves. Prot. Sci. **24**, SI : 95-96 (2015).

23. G. Tartaglia, P. Gaetano, P. Amol S. Campioni, et al. Prediction of aggregation-prone regions in structured proteins. J. Mol. Biol. **380,** 425-436 (2008).

24. Silver D, Huang A, Maddison CJ, et al. (2016) Mastering the game of Go with deep neural networks and tree search. Nature **529**, 484-+.
25. Wada, Junya; Miyazaki, Hiromitsu; Kamada, Rui; et al. Quantitative Correlation between the Protein Expression Level in Escherichia Coli and Thermodynamic Stability of Protein In Vitro. Chem. Lett. **45**, 185-187 (2016).
26. V. T. DeVita, E. DeVita-Raeburn, The Death of Cancer, Farrar, Strauss and Giroux, New York (2015).
27. J. C. Phillips Scaling and self-organized criticality in proteins: Lysozyme $c$. Phys. Rev. E **80**, 051916 (2009).

| Amino acid | βexposed | βHS |
| --- | --- | --- |
| A | 116.65 | 150.3 |
| C | 106.94 | 181.3 |
| D | 253.45 | 63.0 |
| E | 220.67 | 83.5 |
| F | 99.25 | 237.1 |
| G | 142.15 | 126.1 |
| H | 195.57 | 101.9 |
| I | 87.92 | 240.4 |
| K | 221.88 | 106.8 |
| L | 89.94 | 218.9 |
| M | 108.15 | 196.1 |
| N | 195.17 | 88.9 |
| P | 202.45 | 135.8 |
| Q | 207.72 | 92.3 |
| R | 216.62 | 91.9 |
| S | 146.2 | 88.9 |
| T | 133.25 | 144.3 |
| V | 99.65 | 229.3 |
| W | 124.34 | 202.3 |
| Y | 139.72 | 208.2 |

Table I. $\Psi(aa)$ for the β scales here have been adjusted to match the centers and ranges of the KD and MZ hydropathicity scales listed in [24].

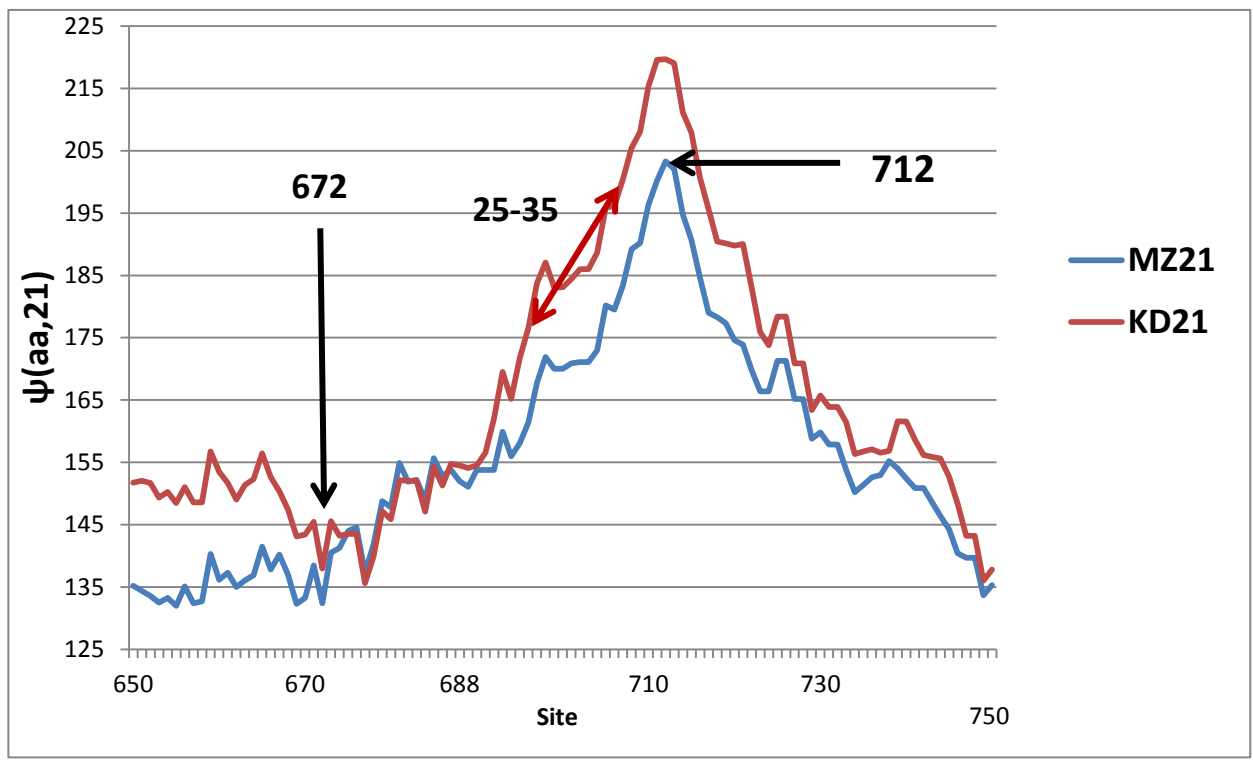

Fig. 1. The amphiphilic character of the 672-712 hydrophobic peak of Aβ is clearest in the hydropathic profile ψ(aa,W) with W = 21, the aa membrane thickness. This choice also identifies the hydrophilic cleavage dip at Asp672, so that the choice W = 21 is overdetermined.

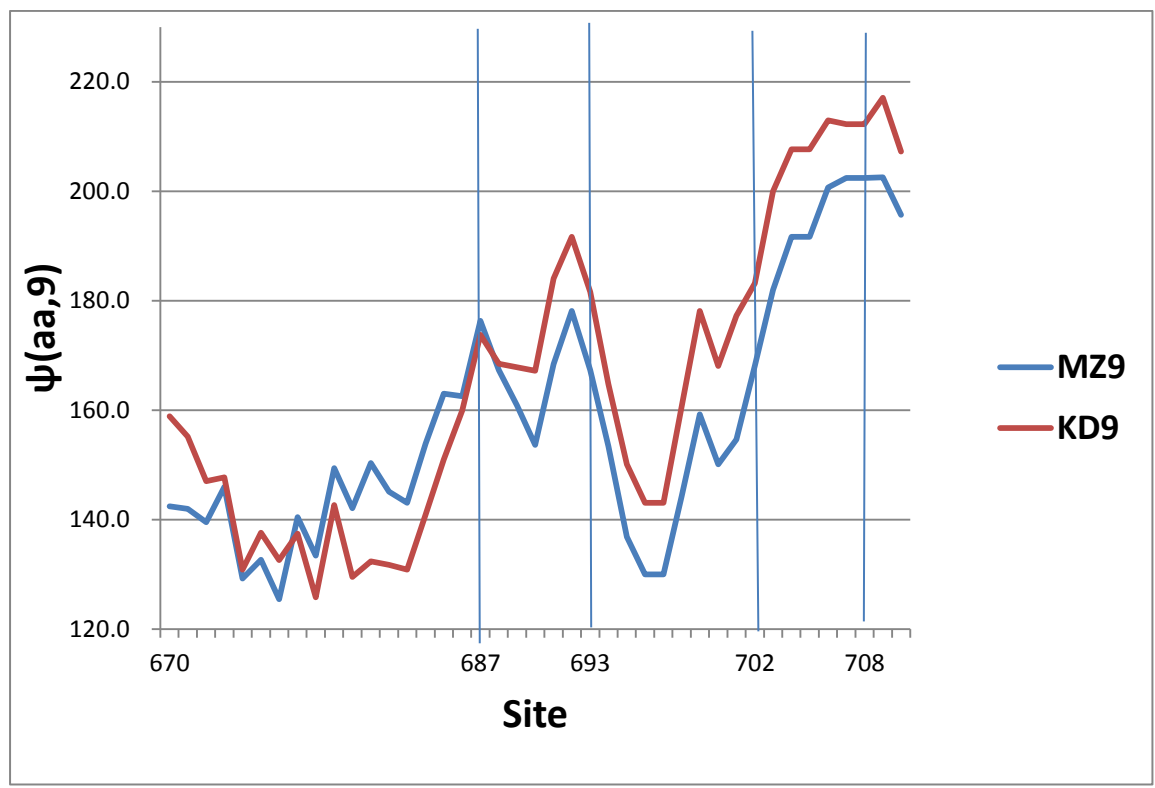

Fig. 2. The smallest Aβ nuclei (7 aa) are 687-693 and 702-708. With W = 9, both nuclei are represented by sharply defined hydrophobic maxima, with the KD scale being slightly sharper.

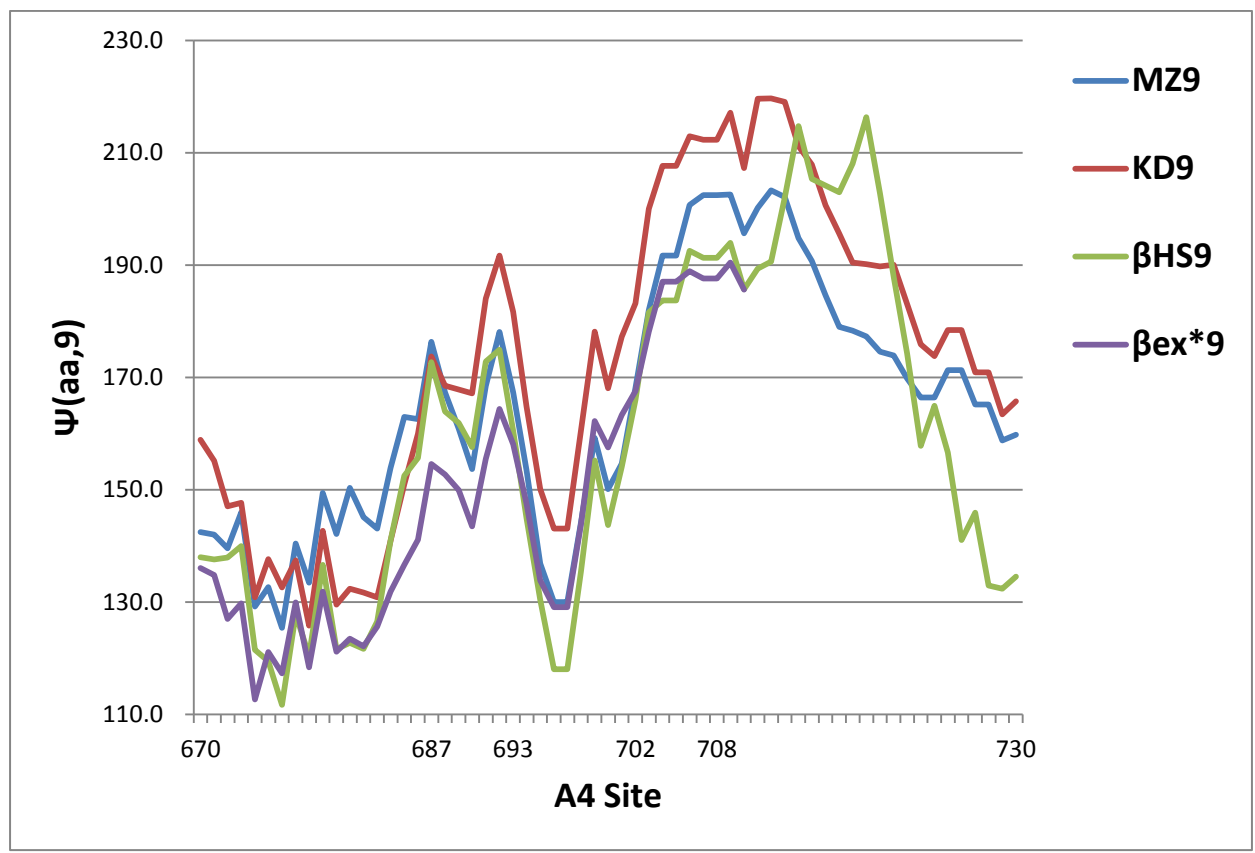

Fig. 3. This complex Aβ figure involves four scales, compared to only the two hydropathic scales in Fig. 1. It also includes two β profiles and uses the small wavelength W = 9 of Fig. 2. Because the FTI scale βexp is anticorrelated with the hydropathic MZ and KD scales, we have plotted instead βexp* = 300 − βexp. The two amyloid nuclei 687-693 (peak) and 702-708 (amphiphilic) are recognized by all four scales. The gap 694-701, which was a false positive in [7] (chiefly because it used W = 6 and multiple adjustable parameters), is strongest in the βHS scale. The βHS scale is based on direct measurements of amyloid formation by central 7 mers 687-693, with the central site 690 mutated through all 20 aa [12].

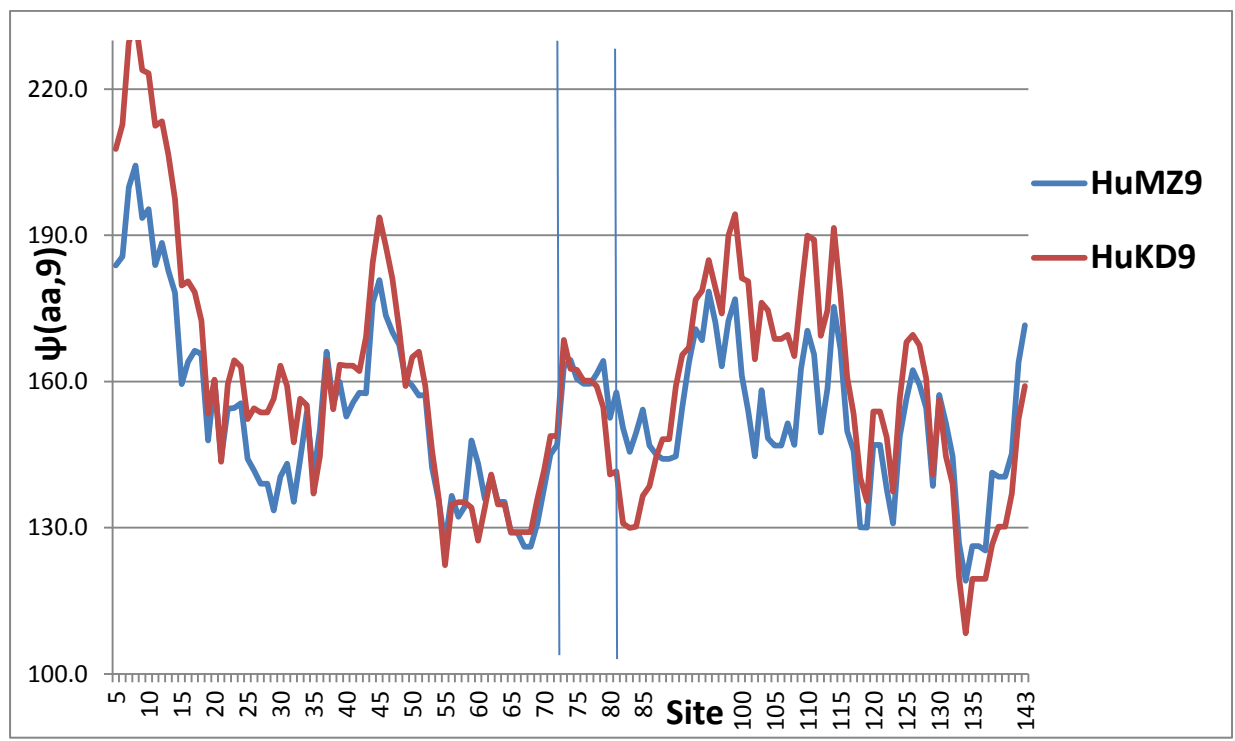

Fig. 4. Hydropathic profiles of human lysozyme *c*, using the long-range MZ scale and the short-range KD scale. The central 7 aa 73-81 amyloid nucleus, indicated by guidelines, is a narrow hydrophobic peak, which is better resolved by the short-range KD scale.

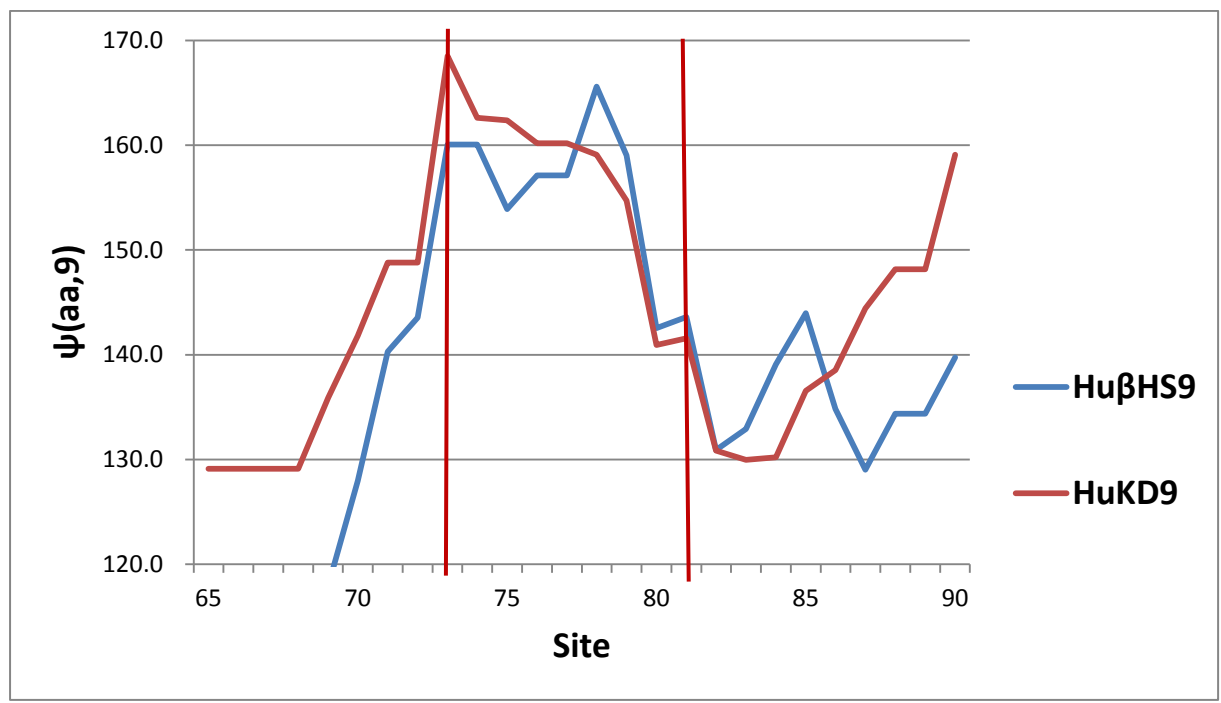

Fig. 5. Enlarged central region of lysozyme in Fig. 4. The N-side of the hydrophobic peak is resolved equally well by both scales, while the C-side is resolved about twice as well by the βHS scale.

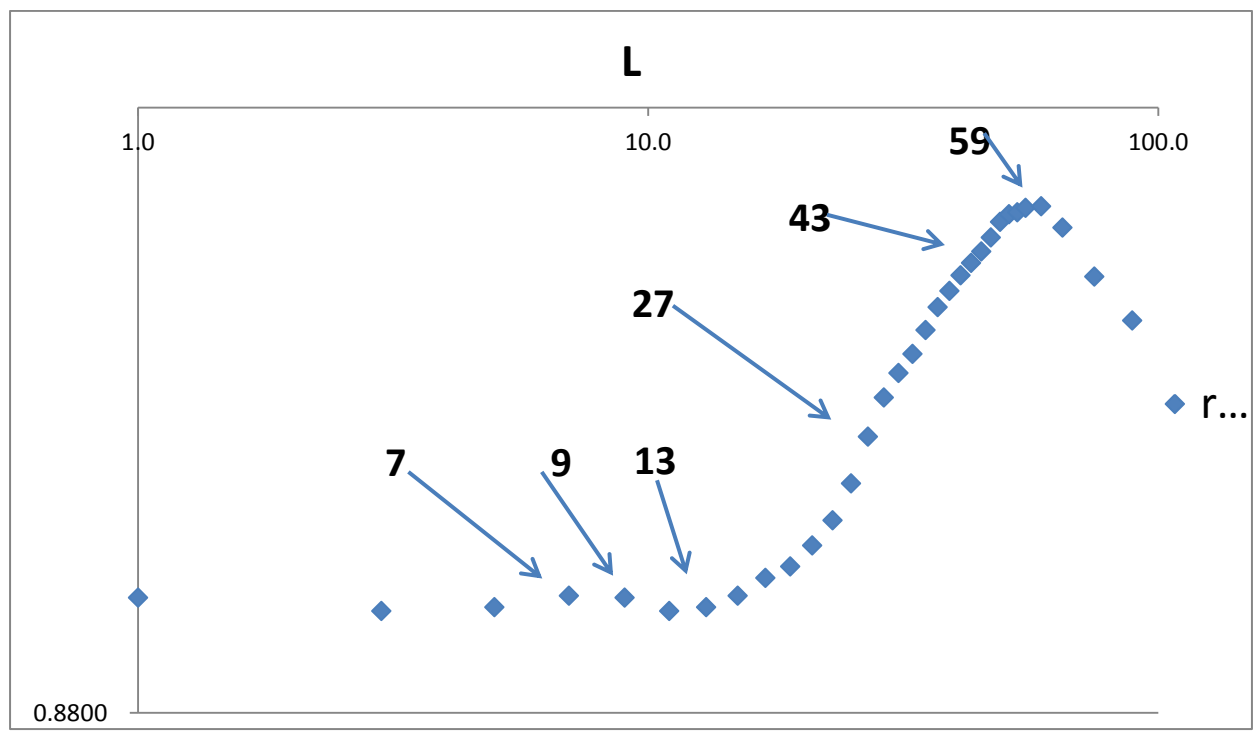

Fig. 6. A log-log plot of the MZ – βHS correlation r(9,L) averaged over all p53. For epitopic lengths L = 7,9 there are weak local maxima at r = 0.893. Above L =13 r climbs superlinearly to L = 27, where a break in slope (a discontinuous inflection point) occurs with r = 0.911. The maximum value r = 0.938 is reached at W = 59. There is a second break in slope at W = 43. The 27-43 linear segment is centered on 35, which is close to the length of 31 aa of the TD 325-355.